\documentclass[conference]{IEEEtran}
\IEEEoverridecommandlockouts

\usepackage{cite}
\usepackage{url}
\usepackage{amsmath,amssymb,amsfonts}
\usepackage{algorithm}
\usepackage{algorithmic}
\usepackage{graphicx}
\usepackage{textcomp}
\usepackage{xcolor}
\usepackage{pifont}

\def\BibTeX{{\rm B\kern-.05em{\sc i\kern-.025em b}\kern-.08em
    T\kern-.1667em\lower.7ex\hbox{E}\kern-.125emX}}
\begin{document}

\title{Trustworthy Provenance for Big Data Science: \\a Modular Architecture Leveraging Blockchain in Federated Settings}



\author{\IEEEauthorblockN{Nicola Giuseppe Marchioro}
\IEEEauthorblockA{\textit{University of Trento}\\
Trento, Italy}
\and
\IEEEauthorblockN{Yannis Velegrakis}
\IEEEauthorblockA{\textit{University of Trento}\\
Trento, Italy}
\and
\IEEEauthorblockN{Valentine Anantharaj}
\IEEEauthorblockA{\textit{Oak Ridge National Laboratory}\\
Oak Ridge, Tennessee, USA}
\and
\IEEEauthorblockN{Ian Foster}
\IEEEauthorblockA{\textit{University of Chicago \& Argonne National Laboratory}\\
Chicago, USA}
\and
\IEEEauthorblockN{Sandro Luigi Fiore}
\IEEEauthorblockA{\textit{University of Trento}\\
Trento, Italy}
}

\maketitle

\begin{abstract}
Ensuring the trustworthiness and long-term verifiability of scientific data is a foundational challenge in the era of data-intensive, collaborative research. Provenance metadata plays a key role in this context, capturing the origin, transformation, and usage of research artifacts. However, existing solutions often fall short when applied to distributed, multi-institutional settings. This paper introduces a modular, domain-agnostic architecture for provenance tracking in federated environments, leveraging permissioned blockchain infrastructure to guarantee integrity, immutability, and auditability. The system supports decentralized interaction, persistent identifiers for artifact traceability, and a provenance versioning model that preserves the history of updates. Designed to interoperate with diverse scientific domains, the architecture promotes transparency, accountability, and reproducibility across organizational boundaries. Ongoing work focuses on validating the system through a distributed prototype and exploring its performance in collaborative settings.

\end{abstract}


\section{Introduction}
As scientific research becomes more data-driven and collaborative, ensuring the reliability and accessibility of research data is crucial to advancing knowledge. Across disciplines, researchers are generating and analyzing vast amounts of data, driving new discoveries and challenges in data collection, management, and integration. However, geographical dispersion, diverse access patterns, and restrictions on data availability create significant challenges in managing and integrating research data effectively \cite{coleman1997building,LEE201574,4804817,app13127082}. In the absence of coordination, these issues can lead to inefficiencies and unpredictability in data retrieval in many real-world scenarios. 

In Open Science \cite{Murray-Rust2008}, FAIR\cite{Wilkinson2016} and TRUST \cite{TRUST} principles have emerged as two essential frameworks, offering structured guidelines to enhance transparency, reliability, trustworthiness and data reuse across the scientific community. These principles have shifted the focus toward making research data more than just accessible; they emphasize structured, interpretable data that can be reused and verified by other researchers. In line with these principles, the significance of unrestricted access to valuable information has become increasingly relevant, as seen in free software movements\cite{stallman2002free} and initiatives that promote the free circulation of information\cite{8760592}. 


A critical barrier to the full achievement of such principles in practical settings is the issue of producing and sharing valuable \textit{provenance information} (i.e., metadata that captures the entire lifecycle of an artifact, providing a complete record from its creation to its consumption). Provenance allows researchers to understand where data originated, how it was processed, and by whom it was created, offering essential context to assess its quality \cite{10.1145/2207676.2208293, 7883515,10.5555/2772879.2773254}. Yet, many current approaches to provenance tracking are insufficient, often lacking the comprehensiveness and transparency required for reliable scientific work. This gap makes it increasingly difficult to verify the authenticity of data, reproduce experiments, or build upon previous findings \cite{doi:10.1161/CIRCRESAHA.114.303819,stodden2014implementing,fetherston2012towards}. These issues can take many forms when provenance is not properly recorded. For instance, version control mistakes can cause outdated or contradictory data to be used \cite{10.1007/11590354_121}. Furthermore, ambiguity regarding who created the data or what software was utilized to manipulate it can raise questions about its validity. Finally, even when provenance is recorded, its utility depends on its accessibility. However, link rot, the phenomenon where data URLs or resources become unavailable, poses a significant risk and can potentially break crucial connections in the research chain \cite{https://doi.org/10.1002/asi.20513,10.1007/978-3-031-27077-2_37}. 

Addressing these challenges requires a multifaceted solution, a provenance management system that strengthens each link in the chain and ensures that all phases of data handling, from generation to distribution, are trustworthy. This system must also be domain-agnostic, flexible enough to integrate with diverse scientific disciplines, and scalable to handle the growing volume of data being produced. By taking advantage of a robust, reliable, and transparent provenance chain, researchers can not only ensure the integrity of their data but also contribute to the wider scientific ecosystem’s need for collaboration and reproducibility.

To this end, this paper introduces a modular, domain-agnostic architecture for provenance tracking in federated environments, leveraging permissioned blockchain infrastructure to guarantee integrity, immutability, and auditability. 

The remainder of this paper is structured as follows:
Section II provides a comprehensive background on provenance metadata and blockchain-based solutions. Section III outlines the methodology, detailing the design decisions that underpin our architecture. Section IV illustrates the operational behavior of the system through real-world use cases, while Section V concludes with a summary of contributions and a discussion of future work.

\section{Background}
Metadata has sometimes been described as ``a love note to the future''\footnote{https://ascii.textfiles.com/archives/3181}, emphasizing its role in preserving the meaning and context of information over time. However, metadata is only useful when it is well-structured and consistently maintained; otherwise, it becomes little more than digital noise, sophistication without any substance. In scientific research, bad metadata can hinder reproducibility, introduce uncertainty, and reduce the long-term utility of data. 

Metadata generally falls into three categories: descriptive metadata, which provides information about the content of a dataset; structural metadata, which describes relationships between different data elements; and administrative metadata, which includes technical details such as file formats and access rights \cite{niso2017metadata}. Provenance metadata spans all three categories: it describes what the data is (descriptive), how it was derived or related to other data (structural), and who modified or accessed it (administrative). This makes provenance a key enabler of transparency and reproducibility in scientific workflows. As research increasingly relies on distributed data sources, computational models, and automated workflows, robust provenance tracking has become a fundamental requirement for scientific integrity. Moreover, provenance aligns closely with the FAIR principles, emphasizing the need for rich metadata to ensure long-term usability and verifiability of research data.

Provenance is well-established in disciplines where data consistency, accountability, and traceability are critical. In databases, provenance tracking is fundamental for query explainability, with models such as why-provenance, how-provenance, and where-provenance \cite{46e660d282474d26ae4965fd1bf8364b,DBS-006} helping users understand how data is derived. In bioinformatics, provenance ensures the reproducibility of genomic analyses and machine learning workflows \cite{info:doi/10.2196/51297,Stevens2007-ms}. Climate Science, through tools like ESMValTool, integrates provenance tracking to ensure transparency in climate model evaluation \cite{gmd-13-1179-2020}.

However, although provenance is established in certain areas, its application within large-scale, distributed scientific infrastructures is fragmented. This lack of standardized, integrated provenance mechanisms raises concerns about metadata completeness, data integrity, and long-term verifiability. The credibility of scientific processes is directly tied to the availability and reliability of provenance records. Without clear provenance, data origin and processing steps remain opaque, making it difficult to assess whether results are reproducible and credible. In short, provenance aims to guarantee:

\begin{itemize}
    \item \textbf{Data authenticity verification: } Ensuring that datasets have not been tampered with.
    \item \textbf{Error tracing and reasoning: } Finding the source of discrepancies or inconsistencies.
    \item \textbf{Accountability: } Unambiguously assigning actions to responsible parties.
\end{itemize}

Transparency in provenance is particularly critical for high-stakes research, such as medical trials, climate modeling, and AI-based decision making, where unreliable data can lead to faulty conclusions with real-world consequences.

Various standards and frameworks have been developed to address provenance tracking and to support the reproducibility of experiments, most notably the \textit{Open Provenance Model}\cite{OPM}, the \textit{W3C Prov}\cite{w3cprov} family, and \textit{RoCrate}\cite{rocrate}.
Although these approaches provide structured provenance tracking, many focus on capturing provenance within isolated experiments or specific workflows, rather than establishing an interconnected ecosystem for end-to-end provenance tracking across research domains. As a result, provenance information often remains fragmented, limiting its effectiveness in long-term data verification and interdisciplinary studies. Furthermore, most existing solutions rely on centralized storage models, where metadata is managed by a single entity.  Centralization is accompanied by inherent risks, such as data loss, incomplete tracking, and even tampering, especially in long-term scientific research where data authentication may be required decades later. These problems have led to a growing interest in decentralized, tamper-proof provenance solutions. 

One promising direction is blockchain-based provenance tracking, which offers an immutable and distributed framework for managing scientific metadata \cite{greenspan2017you,khan2019blockchain,valenta2017comparison,9411380}. However, the choice of blockchain architecture has significant implications for scalability, governance, and security. In public blockchains, such as Bitcoin or Ethereum, anyone can participate in the network, verify transactions, and store data. These blockchains rely on decentralized consensus mechanisms like Proof-of-Work (PoW) or Proof-of-Stake (PoS) to ensure security and prevent tampering \cite{PoW-PoS}. While their transparency and resistance to censorship make them attractive for trustless environments, they come with downsides such as high computational costs, scalability limitations, or dependence on crypto-currencies. 
Conversely, private (or permissioned) blockchains restrict participation to a predefined group of entities, such as research institutions or regulatory bodies. These networks use more efficient consensus mechanisms, significantly reducing overhead while maintaining strong security guarantees. Importantly, private blockchains offer greater control over access permissions, making them better suited for applications where data confidentiality or governance is critical.

When applied to provenance tracking, these differences become crucial. Public blockchains provide an open, verifiable ledger, ensuring transparency and accessibility, which aligns well with open data initiatives and FAIR principles. However, their reliance on energy-intensive mechanisms raises concerns about sustainability, especially for large-scale scientific infrastructures. Additionally, while public ledgers promote openness, they may lack efficient querying mechanisms for metadata retrieval, limiting their practical usability for provenance tracking. On the other hand, private blockchains can use lightweight consensus mechanisms (e.g., Practical Byzantine Fault Tolerance \cite{castro1999practical}), significantly reducing energy consumption while maintaining tamper-proof records. This efficiency makes them particularly well-suited for scientific collaboration, where institutions need trust and interoperability without the computational overhead of public blockchains. 

A federated blockchain model also aligns well with scientific consortia, where multiple organizations participate in provenance tracking without depending on a single point of control. This approach ensures long-term data integrity while allowing controlled access to metadata, which is particularly useful when balancing openness and compliance with data protection regulations. Both \cite{gaetani2017blockchain} and \cite{7973733} propose a decentralized storage system supported by a public blockchain interface used to ensure data integrity and validity. Their strength is that, by using the blockchain as data validator, trust is not blindly placed on an organization anymore, but rather on a cryptographic algorithm that confirms all the operations (PoW). However, as we discussed previously, this choice raises new questions about scalability, governance, and energy efficiency that must be carefully considered.

A more controlled approach consists of adopting permissioned blockchains. Thanks to their nature, permissioned blockchains do not use PoW, but can rely on more sustainable consensus algorithms. Moreover, costs related to the infrastructure are less volatile than in public blockchains, which are oftentimes tied to cryptocurrencies and fluctuating gas fees~\cite{9919978}. This path is explored by Tunstad et al.~\cite{DBLP:journals/corr/abs-1910-05779}, who design a full blockchain-based system to track data provenance, with the primary goal of ensuring the integrity of data-driven research and applications, particularly in distributed environments where data originates from decentralized sources. However, their system lacks a federated, collaborative perspective and does not explicitly address the challenges of data sharing and reuse across organizational boundaries, critical aspects in modern scientific ecosystems guided by FAIR and TRUST principles.

A permissioned blockchain is also used in \cite{demichev2019provenance}, the authors describe a system in which a 2-transaction protocol is executed every time a user wants to interact with the stored data. The first transaction is executed to record the intention of the user, if it gets validated, the operation is performed. Finally, a second transaction records the action taken by the system (potential update of data, data access). In this solution the blockchain acts as a middle layer between the user and the DMS, validating all operations before executing them, in turn enhancing security.
While they propose a blockchain-based approach to manage provenance metadata and access control in semi-distributed environments, their design lacks a federated governance model and does not address the broader scientific needs of reproducibility, metadata evolution, and downstream accountability.

\section{Methodology}
The background research highlights the growing interest in creating provenance information for tracking experiments. Although different approaches have been explored, we still lack a holistic approach that enables trustworthy end-to-end tracking in distributed decentralized environments. To address this issue, we begin by identifying core challenges, which we can define through the following questions:
\begin{enumerate}
    \item How do we keep track of an artifact across disjointed experiments?
    \item How do we ensure that provenance information is consistently trustworthy over time?
    \item How can we ensure that versions of a provenance record are accurately linked and traceable across their lifecycle?
    \item How can the system propagate corrections or retractions to downstream documents or artifacts?
    \item How can we assign accountability for data creation, updates, or misuse in federated environments?
    \item What vulnerabilities arise in provenance tracking systems that rely on centralized or mutable records?
\end{enumerate}
Addressing these is essential to design a provenance system that is not only technically sound but also meaningful and trustworthy in real-world scientific collaboration. 

In the following subsections, we will analyze each question in depth, exploring how its resolution contributes to building a cohesive, robust system.

\subsection{Cross-experiments artifact tracking}
We define an \textit{artifact} as a digital object that participates in a scientific process and is subject to provenance tracking. An artifact may represent data, software, a configuration file, a computational result, or any other digital entity whose origin, transformation, or usage within an experiment is of interest. It is important to note that distinct copies or versions of the same dataset, such as an original and its published replica, are treated as separate artifacts, as they can evolve independently and may be modified without atomic mutual synchronization.

\begin{figure}[h]
    \centering
    \includegraphics[width=0.9\linewidth]{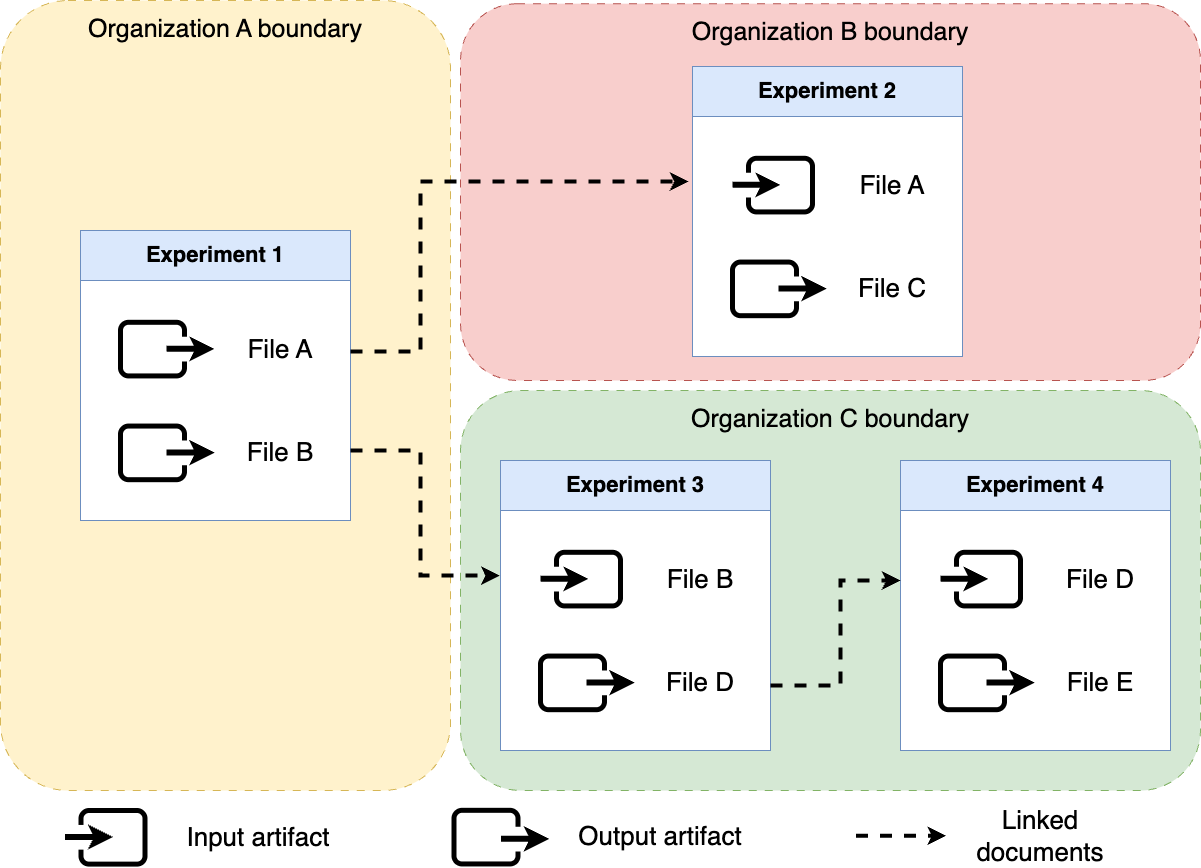}
    \caption{The figure shows linked experiments performed by different organizations. The first experiment (Experiment 1) is performed by Organization~A and generates 2 output files (File A, File B). The generated files are reused respectively as input for Experiment 2 (performed by Organization~B) and Experiment 3 (performed by Organization~C). Finally, the output of Experiment 3 (File D) is in turn used to carry out Experiment 4.}
    \label{fig:linked_exp}
\end{figure}

When a digital object takes part in multiple experiments, it is essential to be able to identify it as the same artifact across different contexts. A common scenario where this is easily applicable is one in which an artifact produced from an experiment \textit{E1} is reused as input for a subsequent experiment \textit{E2}, as seen in Figure \ref{fig:linked_exp}.
In this case, the output of \textit{E1} and the input of \textit{E2} refer to the same resource and should be tracked accordingly to maintain a coherent provenance chain. To this end, we employ Persistent Identifiers (PIDs): globally unique, stable references assigned to digital objects. PIDs are fundamental in our architecture, enabling reliable identification, discovery, and citation of artifacts over time, even if their storage locations or associated metadata evolve. Systems such as DOIs (Digital Object Identifiers), ORCID for author identification, and the Handle System are well-established PID infrastructures. These systems associate each identifier with metadata about the associated object, such as its location, ownership, and version history. 

Our implementation leverages (i) a dedicated prefix for provenance management from the ePIC initiative\footnote{https://www.pidconsortium.net/} coupled with (ii) the Handle Service\footnote{https://www.handle.net/}, a general-purpose PID framework that offers flexibility and long-term stability. By employing this widely adopted technology, we foster integration with existing software stacks and enhance the discoverability of resources. In theory, the binding between an artifact and its PID can be embedded in several places: directly within the artifact itself (if the file format supports it), in the associated provenance record, or even within the PID record metadata itself. Each of these strategies has different implications. Embedding the PID directly into the artifact increases portability but assumes support for metadata embedding, which may not always be available (e.g., in binary formats or proprietary file types). Storing the PID in the provenance document, on the other hand, ensures that the linkage is part of the reproducibility narrative, which is essential for understanding the full lifecycle of the artifact. Lastly, encoding metadata within the PID record can help decouple storage and indexing from the data pipeline, enabling discovery without accessing the original artifact or document.

While these approaches are not mutually exclusive, duplicating information across layers introduces redundancy and requires mechanisms to ensure consistency. We adopt a flexible model that prioritizes the inclusion of PIDs in the provenance document as the authoritative source of linkage, while optionally allowing additional embedding where possible. This accommodates heterogeneous scientific pipelines without forcing a single pattern. To support post-publication improvement of metadata quality, we allow producers to retroactively annotate provenance documents with PIDs, provided that these updates do not alter the recorded sequence of events or original content. This design encourages progressive enhancement of provenance information while preserving its trustworthiness and auditability.

\subsection{Temporal reliability and verification}
Ensuring that provenance information remains verifiable over time introduces multiple concerns, from metadata aging to link degradation and evolving standards. As research artifacts and their descriptions are stored across different platforms and formats, long-term reliability depends on two main factors: (1) the durability of the underlying infrastructure and (2) the immutability and accessibility of the provenance records themselves.

To ensure the consistency and reliability of provenance information over time, our approach centers on immutable and verifiable recording. Resources are committed to a permissioned blockchain, where each transaction represents a time-stamped assertion about an artifact or experimental activity. This guarantees that once a record is written, it cannot be modified or deleted without detection, thus ensuring auditability and tamper-resistance. 
Among the different blockchain technologies explored \cite{valenta2017comparison,9411380}, Hyperledger Fabric (HLF)\footnote{https://www.lfdecentralizedtrust.org/projects/fabric} stood out due to its modular architecture and permissioned nature, making it suitable for a range of applications, including scientific data management. This permissioned model is essential for ensuring privacy, scalability, and performance, key requirements for scientific data systems. HLF’s architecture allows for fine-grained access control, meaning different levels of trust can be granted to different participants based on their roles.

HLF's modular design allows for customization of components such as consensus mechanisms and privacy controls. This modularity ensures that HLF can be tailored to various needs while maintaining high levels of security and trust in the system. In the context of our architecture, this blockchain-backed trust is critical: by maintaining an immutable and cryptographically verifiable log of operations, HLF ensures that each assertion about an artifact—whether it's a data creation event, an update to a provenance document, or an invalidation—is durably recorded. Since each block is linked to its predecessor, tampering with historical data becomes computationally infeasible.
Furthermore, as more research organizations participate in the federated network, the robustness of the system improves. Trust is anchored not in a single entity but in a shared, append-only ledger, where all stakeholders can independently verify the authenticity and history of data and metadata. This guarantees that provenance records remain auditable and reliable over time, reinforcing reproducibility and transparency for external researchers.

\subsection{Architecture design and component interactions}
\begin{figure}[h]
  \centering
  \includegraphics[width=\linewidth]{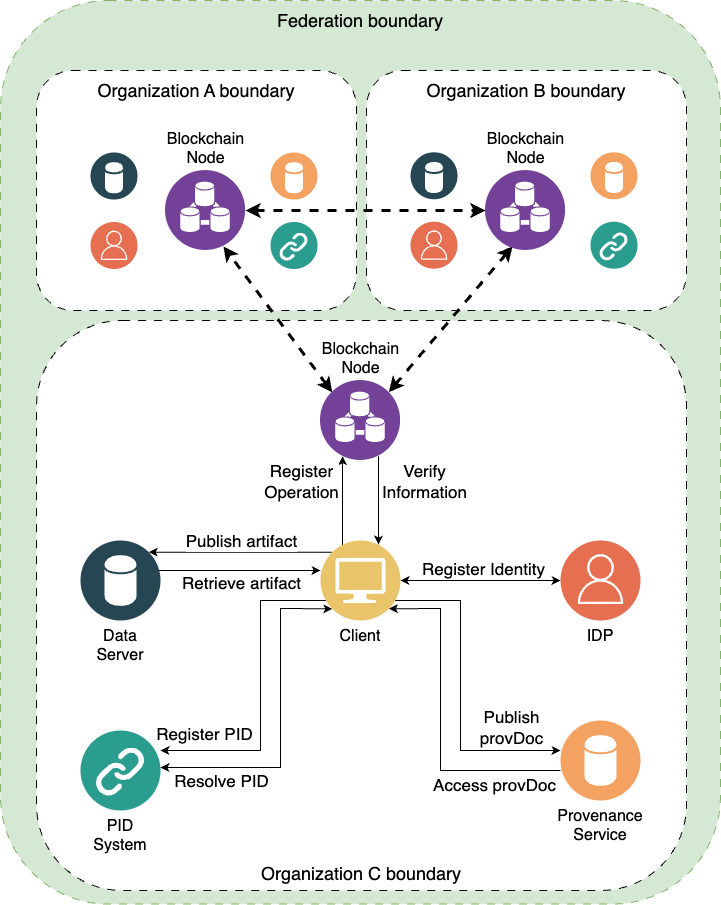}
  \caption{The figure shows the core components of the architecture, the cross-organizational blockchain-based backbone and the set of operations a client can perform at each site. The security infrastructure (green background) implemented at the federation level relies on PKI technology.}
  \label{fig:architecture}
\end{figure}

From the beginning, the architecture was designed to be domain-agnostic, ensuring its applicability across different scientific disciplines. This was achieved by describing the system components in terms of their core role rather than specifying domain-specific functionalities or software. The federated architecture is shown in Figure \ref{fig:architecture} and consists of the following main components:

\begin{itemize}
    \item A client library that allows users to interact with the system and manage provenance data.
    \item A data server responsible for storing datasets and scientific experiment outputs, with the ability to index and search data efficiently.
    \item A (possibly separate) storage system for provenance data, which ensures provenance records are easily accessible and can be searched or linked to corresponding datasets.
    \item A PID service that generates and resolves unique identifiers for both data and provenance records, facilitating easy referencing and citation.
    \item Hashing software used to verify the integrity of data and provenance records, ensuring that any modifications to the files can be detected immediately.
    \item A permissioned blockchain infrastructure, distributed among different organizations, responsible for maintaining the integrity and auditability of both data and provenance information.
\end{itemize}

The client is the primary interface through which users, both producers and consumers of data, interact with the system. Importantly, it also functions as a blockchain client, enabling users to submit and verify provenance records directly on the permissioned ledger.
This direct interaction with the blockchain is critical to the system’s trust model. By allowing the client to communicate with the distributed ledger without relying on an intermediary (such as a proxy or centralized gateway), we ensure that trust is anchored in the technology itself, rather than in a particular server or organizational actor. Users can independently verify the inclusion, integrity, and timestamping of records without having to rely on a trusted third party.
Had the system been designed with a proxy server mediating blockchain access, that proxy would become a central point of failure and potential bottleneck. It would undermine two core principles of the architecture:

Availability: If the proxy server goes offline, users would be unable to submit or verify records, rendering the system partially or completely inaccessible.

Data reliability and integrity: If all interactions pass through a single server, it introduces the risk, whether through compromise or misconfiguration, that the information reaching the blockchain could be manipulated or selectively filtered without user awareness.

By contrast, enabling clients to directly sign and broadcast transactions to the blockchain ensures that all submissions are cryptographically attributable to their originators and that no trusted intermediary is needed to validate the authenticity or timing of a record. This decentralized interaction pattern reinforces end-to-end verifiability, a crucial feature for scientific environments where reproducibility, auditability, and transparency are essential.
To enable secure and verifiable interactions with the blockchain, each client must possess a cryptographic identity recognized by the Hyperledger Fabric network. This identity is signed by a Certificate Authority (CA) that is trusted by the blockchain network. Access control is enforced by access control lists (ACLs) and endorsement policies, which determine what operations a user can perform based on their organization and role. Authentication is handled via public-key infrastructure (PKI), which means that each user possesses a private key and a corresponding certificate (public key). Authorization is managed at the network level, so only users from trusted organizations are allowed to execute certain types of transactions. Data producers and curators are assigned to organizations participating in the federation (such as research laboratories or institutions), and only users within these organizations are authorized to publish to the blockchain. In contrast, consumers who only need to verify or query provenance records are assigned to a fictional, read-only organization. This organization issues certificates that allow users to perform queries without granting write permissions.
To make this model work, the client application manages identity at the user level. During registration, the client communicates with an identity provider (IDP) to register the user and obtains a private/public key pair. It submits the public key to the HLF network, where it is registered as a new identity under the appropriate organization. Once the network approves this identity, the user can begin interacting with the blockchain.

\begin{table}[h]
    \centering
    \caption{Enabled CRUD operations in the system. 'Deletion' is implemented as invalidation to preserve provenance integrity and support explainability.}
    \begin{tabular}{|c|c|c|}
        \hline
        Operation & Artifact & Provenance doc \\ \hline
        \textbf{C}reate & \ding{51} & \ding{51} \\ \hline
        \textbf{R}ead & \ding{51} & \ding{51} \\ \hline
        \textbf{U}pdate & \ding{55} & \ding{51} \\ \hline
        \textbf{D}elete & \ding{51}* & \ding{55} \\ \hline
    \end{tabular}
    \label{tab:op_tables}
\end{table}
The client supports the operations listed in Table \ref{tab:op_tables}. Notably, updates to artifacts are not permitted, as allowing modifications could lead to inconsistencies and potentially enable result falsification. Likewise, provenance records cannot be deleted to prevent the possibility of retroactively hiding published information. In the same vein, artifact deletion is implemented as invalidation rather than actual removal, preserving referential integrity and ensuring consistent system behavior. The invalidation mechanism can be extended by implementing custom policies for handling derivative products of an invalidated artifact. For example, when an artifact is marked as invalid, the system can automatically trigger review workflows or propagate invalidation notices to all dependent records. This ensures that downstream artifacts remain consistent with the status of their sources and helps maintain the overall integrity of the provenance graph. Finally, it is important to notice that, to prevent falsification, updates to provenance files are only possible as long as the system guarantees the storage of the old provenance record. Versioning of provenance file is achieved through the use of the PID record. When a new version of a provenance record is published, the PID record of the previous version is updated to contain a reference to the new version. The publication operation is then registered on the blockchain.  This pattern provides both ease of access and strong integrity guarantees. Since versioning information is directly embedded within the PID record, users and automated systems can easily retrieve the complete version history of a provenance file. At the same time, the blockchain ensures that all operations are transparently and tamper-proofly logged, preserving trust in the system’s data lineage.
To illustrate how the system handles provenance record updates, we present a sequence diagram that highlights the coordinated interaction among the client, blockchain, PID resolver, and provenance manager.
The following preconditions must occur before the operation can be performed:
\begin{enumerate}
    \item A client is configured with a valid identity for interacting with the HFL network
    \item A provenance record has already been published and is resolvable via a PID.
    \item The identity has the right to update the provenance record, either because it was initially created by it, or has been authorized.
    \item A new version of the provenance record was generated and the user has it in his local machine. This version enriches the previous one by adding more details regarding one aspect of the experiment.
    \item The provenance manager is a trusted component that allows atomic update of a provenance record.
\end{enumerate}

Figure \ref{fig:UpdatePROV_seq} summarizes the interactions necessary to perform this update:

\begin{figure*}[h]
  \centering
  \includegraphics[width=\linewidth]{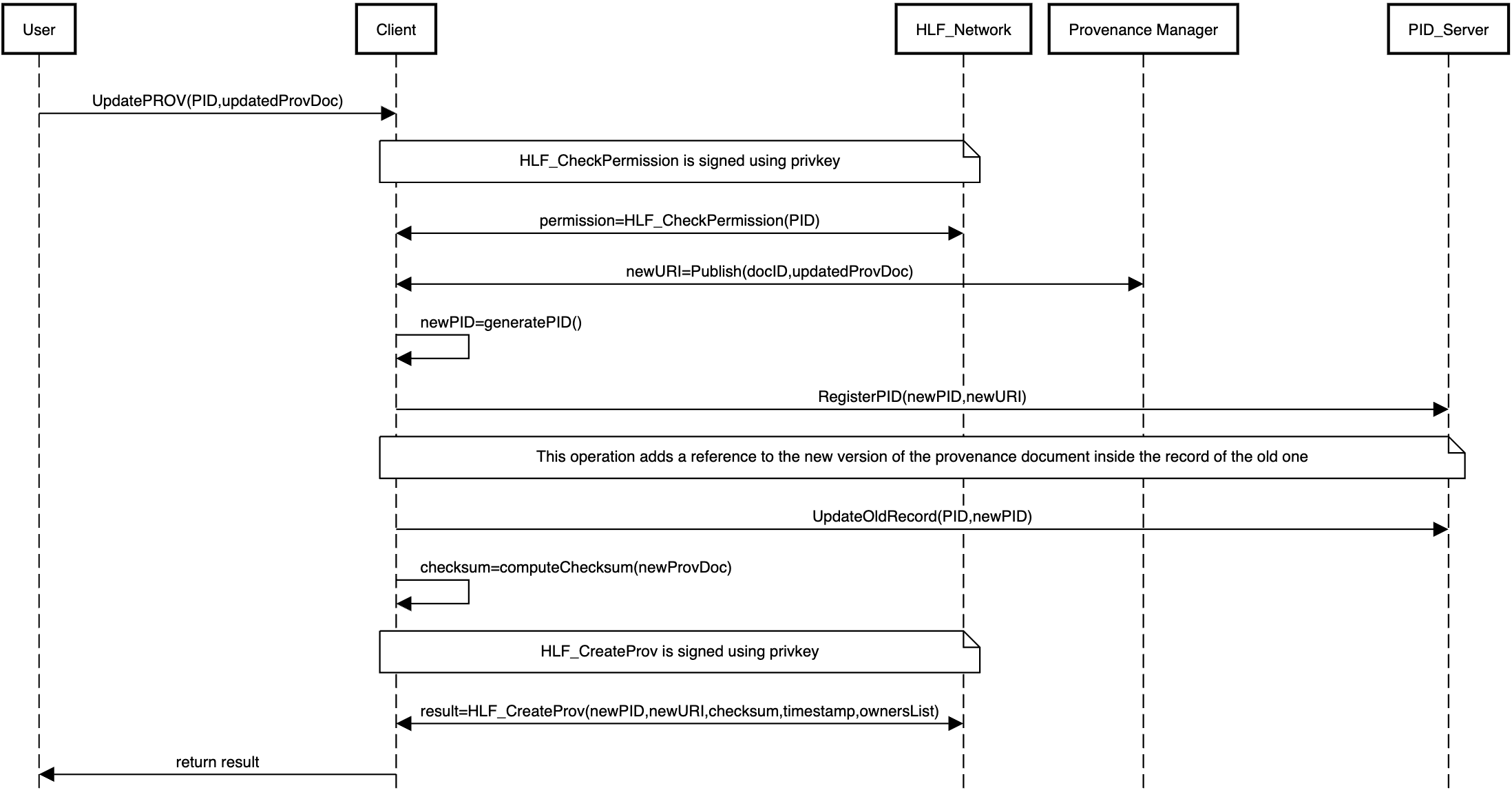}
  \caption{Sequence diagram for the \textit{update}  of a provenance record. This operation allows enriching the content of an existing provenance document. By also maintaining a copy of the previous version (and cross-referencing it with the updated one) users can easily understand the motivation behind the update, or dispute it, if the update was improperly performed.}
  \label{fig:UpdatePROV_seq}
\end{figure*}

Algorithm \ref{alg:HLFupdate} outlines the logic of the HLF\_UpdateProv chaincode function used to apply the update. In HLF, such operations are executed via invoke transactions, where the client submits a signed proposal to the endorsing peers of the relevant organization(s). These peers simulate the transaction, return signed endorsements, and upon satisfying the endorsement policy, the transaction is submitted to the ordering service and subsequently committed to the ledger by all peers.
\begin{algorithm}
\caption{Update Provenance Record}
\begin{algorithmic}[1]
    \STATE \textbf{HLF\_UpdateProv}($PID$, $newURI$, $newChecksum$, $caller$, $timestamp$, $permission$) 
        \IF{$\neg$ checkAuth($PID$,$permission$)}
            \RETURN \texttt{"Error: Unauthorized user"}
        \ENDIF
        \STATE value $\gets$ GetState($PID$)
        \IF{value = null}
            \RETURN \texttt{"Error: Resource not found"}
        \ELSE
        \STATE ($URI$, $checksum$, $version$, $ownersList$, $oldTimestamp$) $\gets$ value
            \STATE newVersion $\gets$ version + 1
            \STATE PutState($PID$, ($newURI$, $newChecksum$, $newVersion$, $ownersList$, $timestamp$))
            \RETURN \texttt{"Success: Resource updated successfully"}
        \ENDIF
\end{algorithmic}
\label{alg:HLFupdate}
\end{algorithm}


\section{Scientific Use Case Patterns}

To illustrate how the proposed system supports real-world scientific workflows, we present five representative use cases (UCs) that demonstrate its application across typical scientific research scenarios. 

All of them are under development and have been preliminarily demonstrated in a distributed setting at the High Performance Climate Informatics Laboratory of the University of Trento to assess the feasibility and versatility of the proposed approach across diverse use cases. Some of them, such as UC1, are being developed into an European Open Science Cloud (EOSC) testbed as part of the Horizon Europe EOSC-Beyond project\footnote{https://www.eosc-beyond.eu/}, with a focus on federated provenance management; others, like UC2, will be demonstrated in a trans-Atlantic testbed involving the University of Trento, Oak Ridge National Laboratory (ORNL) and Argonne National Laboratory (ANL) as part of an international research effort on large-scale federated climate data management. 

Though all of them are presented in the following sub-sections, for page limit constraints, the implementation details of the different use cases, as well as experimental benchmarks, are not covered in this paper and will be discussed in detail in a future work.

\subsection{UC1: End-to-End Provenance in Climate Data and AI Forecasting}
A climate researcher (data producer) publishes a dataset measuring a specific climate variable (e.g., soil moisture or surface temperature) across multiple geographic locations. The dataset is registered in the system with a PID and linked to a provenance record detailing its origin, preprocessing, and sensor information.

A second researcher uses this dataset to train a machine learning model to perform spatial interpolation or forecasting. The resulting model, once validated, is used to generate visual forecast maps (images) over unmapped or data-scarce areas.

These forecast images are included in a scientific publication. The figure caption contains a PID pointing to the output artifact. Thanks to the system's provenance support, anyone reading the paper can trace the image back through the model, to the training dataset, and all the way to the original measurements, establishing a verifiable data lineage.

\subsection{UC2: Propagation of Data Corrections and Downstream Invalidation}
A dataset previously published within a federated data archive (i.e., the Earth System Grid Federation (ESGF) \cite{CINQUINI2014400} in the climate change domain) is later retracted due to an error, with the correction communicated via a proper service (i.e., the ESGF Errata Service \cite{gmd-14-629-2021}).

Because this dataset was previously used to train models or generate downstream results, invalidating the primary data triggers a cascade. Through the system, it is possible to identify all derivative artifacts and automatically flag them as affected. This helps consumers avoid using outdated or invalid results and promotes responsible reanalysis or retraction where needed.
\subsection{UC3: Iterative Model Development and Provenance Tracking}
A team is developing a predictive model for drought forecasting. During development, they iteratively refine their model architecture and retrain with new versions of the input data. Each new iteration, whether it’s a model checkpoint, a new training dataset version, or a parameter sweep—is published as a separate artifact with a unique PID and a provenance record linked to previous versions.

Consumers of the model (e.g., another research team or a policy agency) can inspect the full history of the model's evolution. They can choose to use the latest version or a specific iteration that fits their validation needs. This transparency in the training phase fosters confidence in the credibility of the model and supports reproducibility.

\subsection{UC4: Post-Publication Artifact Linking and PID Enrichment}
After the publication of an experiment, a previously unavailable data artifact (e.g., a code repository, supplemental dataset, or visualization output) is made accessible and assigned a PID. To reflect this update, a new version of the provenance record is issued. The structure of the record remains unchanged, but the entity representing the artifact is enriched with a reference to the PID, enabling direct access. This non-invasive enrichment improves discoverability and usability without altering the original context or compromising the traceability of the published experiment.

\subsection{UC5: Retrospective Detail Enhancement via Activity Decomposition}
A researcher reviews a previously published workflow to provide more granular documentation of how a key step was executed. The original activity, previously described as a single black-box process, is now decomposed into multiple sub-activities, each with its own metadata and links to intermediate artifacts. A new version of the provenance record is published, replacing the original activity with this expanded structure. The updated record maintains links to prior versions via the PID mechanism, preserving the chain of updates while enhancing scientific process transparency and results interpretability.

\section{Conclusions}
In this work, we have introduced a modular and domain-agnostic architecture for trustworthy provenance tracking in federated scientific settings. By leveraging a permissioned blockchain and a well-defined interaction model between core components, the framework ensures transparency, immutability, and accountability in data workflows. Its design deliberately abstracts from any specific domain, enabling seamless integration into a wide range of scientific research infrastructures.

Through a combination of persistent identifiers, verifiable versioning, and decentralized trust anchoring, the system offers a robust foundation for provenance tracking that aligns with the FAIR and TRUST principles. It supports reproducibility, error tracing, and data integrity without introducing centralized points of failure, thus reinforcing the credibility of collaborative scientific research.

To validate and explore these benefits in real-world settings, we are implementing a set of scientific use cases across multiple organizations in the framework of EU-funded projects as well as in a  trans-Atlantic testbed focused on federated climate data management. This activity will also serve as a foundation to quantitatively assess performance characteristics under different federation sizes and usage patterns. In that respect, our goal is to better understand the trade-offs involved in adopting blockchain-backed provenance and to guide its optimization for scalable and sustainable scientific use. 

\section*{Acknowledgments}

This work was partially funded by the EU Horizon Europe interTwin project (GA 101058386) and the EU Horizon Europe EOSC-Beyond project (GA 101131875). Moreover, this work was also partially funded under the NRRP, Mission 4 Component 2 Investment 1.4, by the European Union – NextGenerationEU (proj. nr. CN\_00000013). This research used resources of the Oak Ridge Leadership Computing Facility at the Oak Ridge National Laboratory, which is supported by the Office of Science of the U.S. Department of Energy under Contract No. DE-AC05-00OR22725.

\bibliographystyle{plain}
\bibliography{bibliography}
\vspace{12pt}

\end{document}